\documentstyle[aps,epsfig]{revtex}

\newcommand{\bea}{\begin{eqnarray}}
\newcommand{\eea}{\end{eqnarray}}
\newcommand{\be}{\begin{equation}}
\newcommand{\ee}{\end{equation}}

\newcommand{\gsim}{\mathrel{\hbox{\rlap{\lower.55ex \hbox {$\sim$}}
                   \kern-.3em \raise.4ex \hbox{$>$}}}}
\newcommand{\lsim}{\mathrel{\hbox{\rlap{\lower.55ex \hbox {$\sim$}}
                   \kern-.3em \raise.4ex \hbox{$<$}}}}

\def\roughly#1{\mathrel{\raise.3ex\hbox{$#1$\kern-.75em%
\lower1ex\hbox{$\sim$}}}}
\def\lsim{\roughly<}
\def\gsim{\roughly>}

\begin{document}

\twocolumn[\hsize\textwidth\columnwidth\hsize\csname @twocolumnfalse\endcsname

\title{Implications of the ALEPH $\tau$-Lepton Decay Data 
      for Perturbative and Non-Perturbative QCD}

\author{Thomas Sch\"afer$^{1,2}$ and Edward V. Shuryak$^1$}

\address{$^1$ Department of Physics and Astronomy, 
     State University of New York, 
     Stony Brook, NY 11794-3800 \\ 
     $^2$ Riken-BNL Research Center, Brookhaven National 
     Laboratory, Upton, NY 11973}


\maketitle    
\begin{abstract}
 We use ALEPH data on hadronic $\tau$ decays in order to calculate
Euclidean coordinate space correlation functions in the vector 
and axial-vector channels. The linear combination $V-A$ receives
no perturbative contribution and is quantitatively reproduced by 
the instanton liquid model. In the case of $V+A$ the instanton
calculation is in good agreement with the data once perturbative
corrections are included. These corrections clearly show the 
evolution of $\alpha_s$. 
We also analyze the range of validity of the Operator Product 
Expansion (OPE). In the $V-A$ channel we find a dimension $d=6$ 
contribution which is comparable to the original SVZ estimate, but 
the instanton model provides a different non-singular term of the 
same magnitude. In the $V+A$ case both the OPE and the instanton 
model predict the same $d=4$ power correction induced by the 
gluon condensate, but it is masked by much larger perturbative
contributions. We conclude that the range of validity of 
the OPE is limited to $x\lsim0.3$ fm, whereas the instanton
model describes the data over the entire range.

\end{abstract}
\vspace{0.1in}
]
\begin{narrowtext}   

\newpage


 1. Quantitative understanding of the interface between 
perturbative and non-perturbative effects is the central 
problem in QCD dynamics. Historically, QCD sum rules based 
on the Operator Product Expansion (OPE) 
\cite{SVZ,Shifman:1998rb,Shifman:2000jv} constituted
the first serious attempt to describe non-perturbative phenomena 
in QCD. The initial application of QCD sum rules to vector
and axial vector meson lead to very promising results. 
It was soon discovered, however, that not all hadrons are
alike \cite{Novikov:1981xj}. Phenomenology demands that 
non-perturbative effects in scalar and pseudo-scalar 
channels, both meson and glueball, are much bigger than 
in the vector channels. This fact is not reproduced by 
the OPE but it was realized that direct instanton effects
appear in exactly those channels in which non-perturbative
effects are large. This observation gave rise to the instanton 
liquid model \cite{Shuryak:1982ff}. 

  The available information on hadronic correlation functions, 
both from experimental data, the OPE and other exact results was 
reviewed in \cite{Shuryak:1993kg}. Since then, the high statistics 
measurement of hadronic $\tau$ decays $\tau\rightarrow \nu_\tau
+{\rm hadrons}$ by the ALEPH experiment at CERN \cite{aleph1}
has significantly improved the experimental situation in the 
vector and axial-vector channel. The purpose of this paper is 
to compare these results with theoretical predictions, both
from the OPE and instanton models. In particular, we would
like to assess the range of applicability of the two approaches
and put improved constraints on the parameters that enter
into the calculations. Translating the spectral functions 
measured by the ALEPH collaboration into Euclidean coordinate 
space correlation functions will also allow precise comparison 
of the experimental data with improved lattice calculations 
along the lines of \cite{Chu:1993mn}.


2. In the following, we shall consider the vector and 
axial-vector correlation functions $\Pi_V(x) = \langle 
j^a_\mu(x)j^a_\mu(0)\rangle$ and $\Pi_A(x)=\langle
j^{5\, a}_\mu(x) j^{5\, a}_\mu(0)\rangle$. Here, $j^a_\mu(x)
=\bar{q}\gamma_\mu\frac{\tau^a}{2}q$, $j^{5\,a}_\mu(x)
=\bar{q}\gamma_\mu\gamma_5\frac{\tau^a}{2}q$ are the 
isotriplet vector and axial-vector currents. The 
correlation functions have the spectral representation 
\cite{Shuryak:1993kg}
\be
 \Pi_{V,A}(x) = \int ds\, \rho_{V,A}(s)D(\sqrt{s},x),
\ee
where $D(m,x) = m/(4\pi^2 x)K_1(mx)$ is the Euclidean 
coordinate space propagator of a scalar particle with 
mass $m$. We shall focus on the linear combinations
$\Pi_V+\Pi_A$ and $\Pi_V-\Pi_A$. These combinations 
allow for a clearer separation of different 
non-perturbative effects. The corresponding spectral
functions $\rho_V\pm\rho_A$ which are measured by the 
ALEPH collaboration are shown in Fig. \ref{fig_data}. 


%

 In QCD, the vector and axial-vector spectral functions have to
satisfy chiral sum rules. If we assume that $\rho_V(s)-\rho_A(s)=0$ 
above the maximum invariant mass $s=m_\tau^2$ for which the spectral 
functions can be measured, then we find that the ALEPH data satisfy all
chiral sum rules within the experimental uncertainty. However, the 
central values of the sum rules differ significantly from the chiral 
predictions \cite{aleph1}. In general, both $\rho_V$ and $\rho_A$
are expected to show oscillations of decreasing amplitude 
\cite{Shifman:1998rb}. If we set $\rho_V(s)-\rho_A(s)=0$ above
an arbitrarily chosen invariant mass $s_0$ this will lead to the
appearance of spurious dimension $d=2,4$ operators in the correlation 
functions at small $x$. For this reason we have decided to use
$s_0=2.5\,{\rm GeV}^2$, which is slightly below the tau mass
but allows all chiral sum rules to be satisfied. The reader 
should be aware of the fact that we have, in effect, slightly 
moved the data points in the small $x$ region within the error 
bars reported by the ALEPH collaboration. Finally we add the 
pion pole contribution, which is not shown in Fig. \ref{fig_data},
to the axial vector spectral function. This corresponds to 
an extra term $\Pi_A^\pi(x) = f_\pi^2m_\pi^2 D(m_\pi,x)$. The 
resulting correlation functions $\Pi_V(x)\pm\Pi_A(x)$ are shown 
in Fig. \ref{fig_cor1}.


3. We begin our analysis with the combination $\Pi_V-\Pi_A$. This 
combination is sensitive to chiral symmetry breaking, while 
perturbative diagrams, as well as gluonic operators cancel.

\begin{figure}[tbh]
\begin{center}
\leavevmode
\epsfig{file=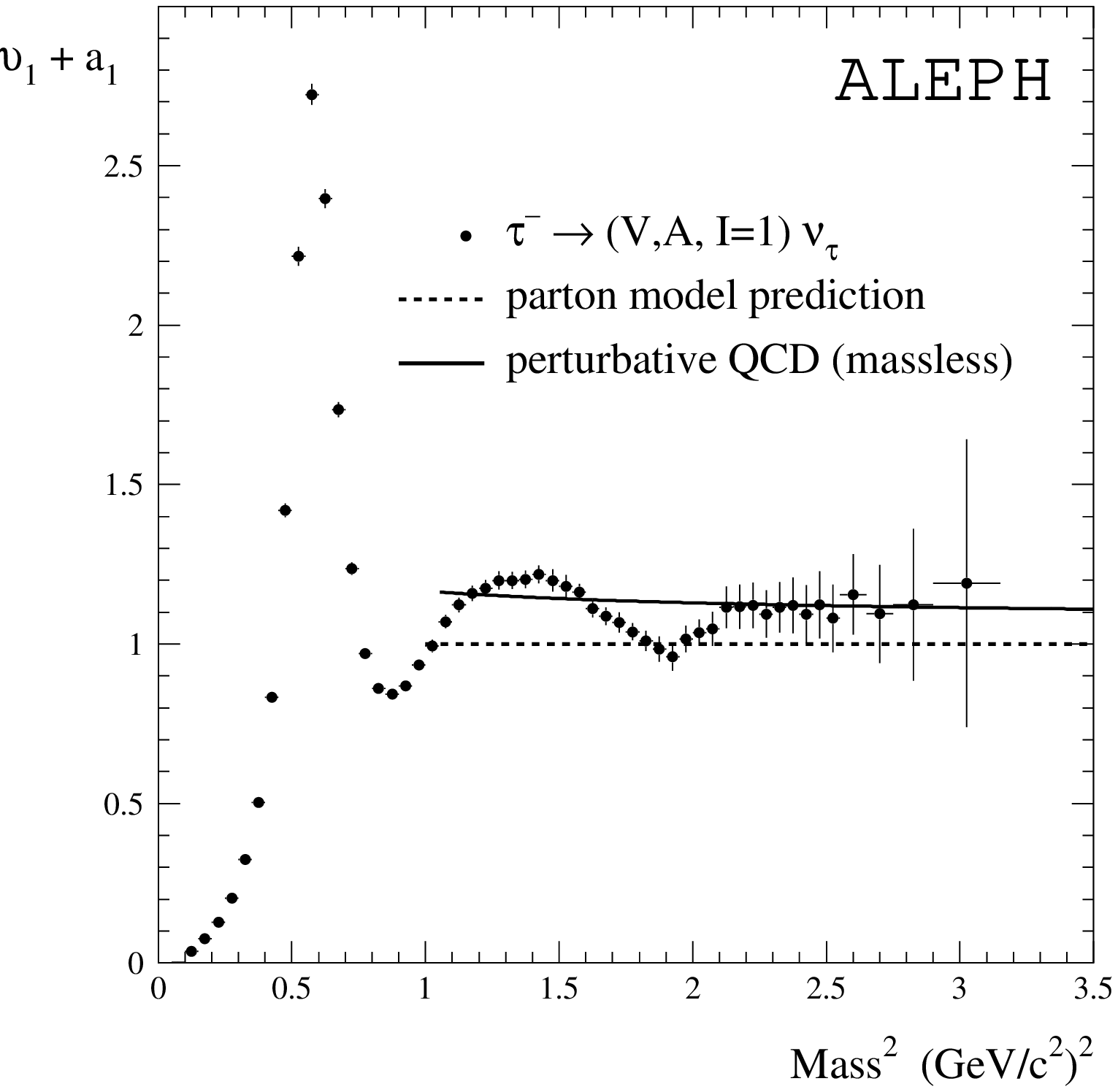,width=6.5cm,angle=0}
\epsfig{file=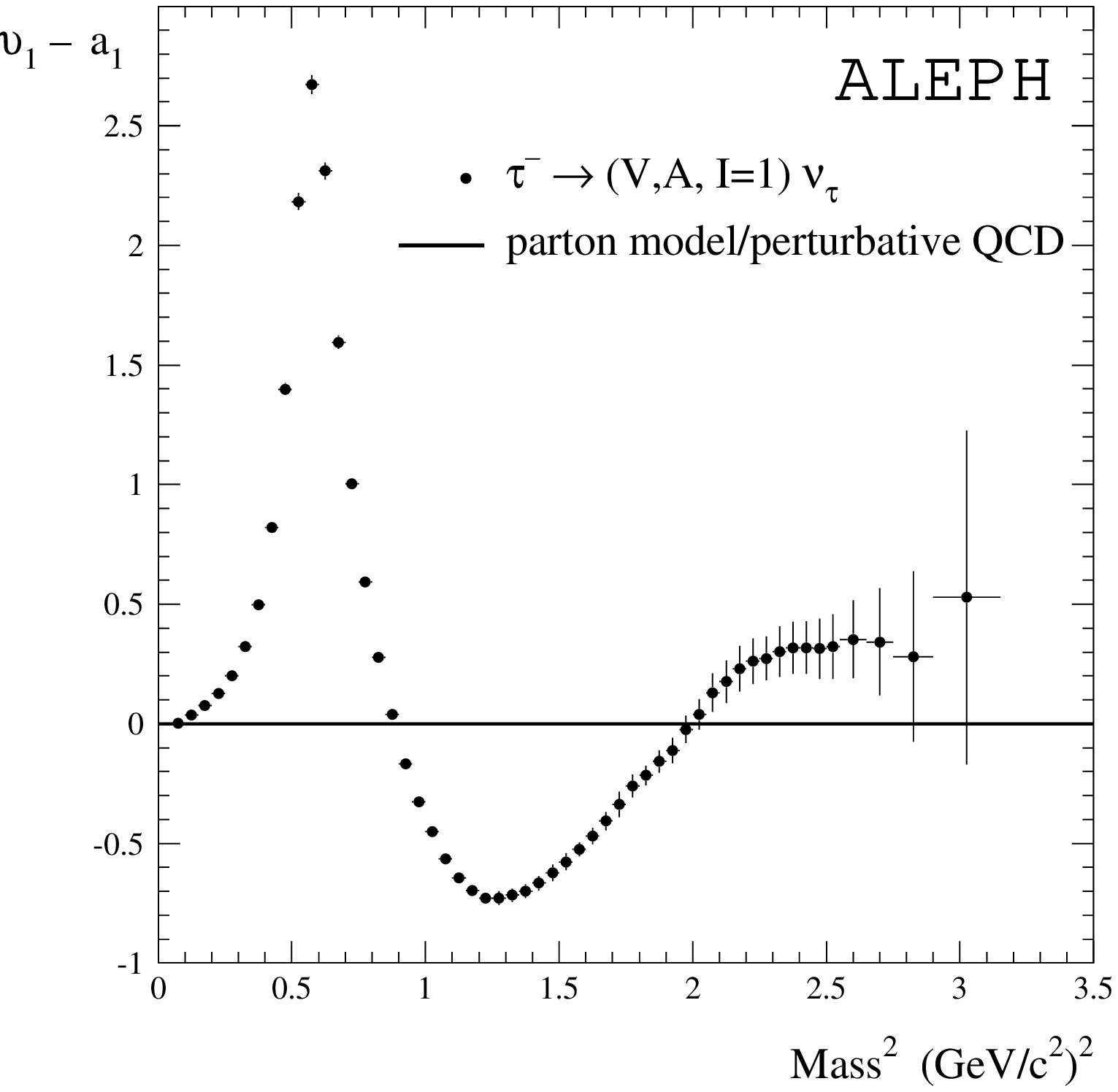,width=6.5cm,angle=0}
\end{center}
\caption{\label{fig_data}
Spectral functions $v(s)\pm a(s)$ $=$ $4\pi^2(\rho_V(s)+\rho_A(s))$
extracted by the ALEPH collaboration.}
\end{figure}

 In Fig. \ref{fig_cor1} we compare the measured correlation
functions with predictions from the instanton model. These
predictions are described in great detail in
\cite{Shuryak:1993ke,Schafer:1996uz} and the review 
\cite{Schafer:1998wv}. The main assumption
is that the QCD vacuum is dominated by strong non-perturbative
field configurations, instantons. In the simplest model, the
random instanton liquid (RILM), the instanton positions and 
color orientations are distributed randomly. The ensemble is 
characterized by two numbers, the instanton-anti-instanton 
density $(N/V)=1 \,{\rm fm}^{-4}$ and the average instanton
size $\rho=1/3$ fm. These parameters were fixed a long time 
ago using the requirement that they must reproduce the 
phenomenological values of the quark and gluon condensates 
\cite{Shuryak:1982ff}.
The agreement of the instanton prediction with the measured
$V-A$ correlation is impressive and extends all the way from
short to large distances. At distances $x>1.25$ fm both
combinations are dominated by the pion contribution
while at intermediate $x$ the $\rho,\rho'$ and $a_1$
resonances contribute. 

 In order to study the validity of the operator product expansion
we have to study the short distance region in more detail. The OPE 
predicts that the $V-A$ correlation function starts with the 
the following quark-anti-quark operators of dimension $d=4$ 
and $d=6$
\bea
\frac{\Pi_V(x)-\Pi_A(x)}{2\Pi_0(x)} &=& 
  -\frac{\pi^2}{4}m\langle\bar{q}q\rangle  x^4 \nonumber \\
 & & \mbox{}
  +\frac{\pi^3 }{9}\alpha_s(x)\langle{\bar{q}q}\rangle^2
   \log(x^2)x^6+\ldots .
\eea
The value of the dimension $d=4$ operator is determined by 
the Gell-Mann-Oakes-Renner (GMOR) relation to be $(x/1.66\,
{\rm fm})^4$. Using $\langle\bar{q}q\rangle=-(230\,{\rm MeV})^3$ 
and the one-loop running coupling constant we also estimate the 
$d=6$ operator as $(x/0.66\,{\rm fm})^6$. This implies that the 
$d=6$ operator totally dominates over the $d=4$ operator.


 This estimate can be checked by considering the value of the 
$d=6$ operator as a free parameter and trying to extract it
from the measured data. Because the $d=4$ operator is so 
small we use the GMOR value. A similar determination of
power corrections was already done by the ALEPH collaboration
using moments sum rules \cite{aleph1,Braaten:1992qm,Ioffe:2000}. 
Nevertheless, fitting the OPE coefficients in coordinate space 
provides important additional insight. The results depend on the 
coordinate range $[0,x_{m}]$ used in the fit, but for $x_{m}<0.3$ 
fm this dependence is weak. The result for $x_{m}=0.3$ fm, also shown 
in Figs. \ref{fig_cor1},\ref{fig_cor2}, is
\be
\frac{\Pi_V(x)-\Pi_A(x)}{2\Pi_0(x)} = 
  \left(\frac{x}{1.66\,{\rm fm}}\right)^4
+ \left(\frac{x}{0.66\,{\rm fm}}\right)^6 + \ldots .
\ee
We find that the size of the dimension $d=6$ term 
agrees with the SVZ prediction. However, the range of 
convergence of the OPE is only $x\lsim 0.3$ fm. We 
also note that the accuracy of the data 
in this regime is very poor. 

\begin{figure}[tbh]
\begin{center}
\leavevmode
\vspace*{-1.2cm}
\epsfig{file=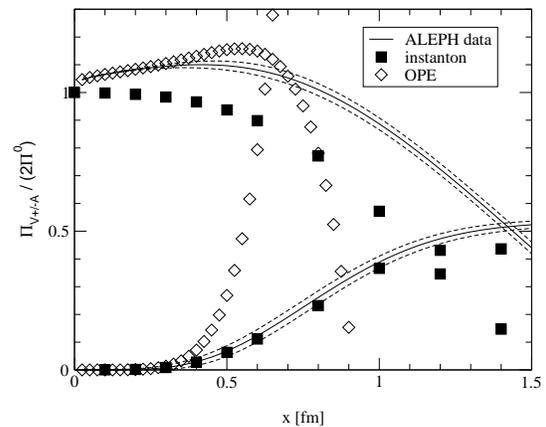,width=8cm,angle=-90}
\end{center}
\vspace*{-1.0cm}
\caption{\label{fig_cor1}
Euclidean coordinate space correlation functions $\Pi_V(x)
\pm \Pi_A(x)$ normalized to free field behavior. The solid lines
show the correlation functions reconstructed from the ALEPH spectral
functions and the dotted lines are the corresponding error band.
The squares show the result of a random instanton liquid model
and the diamonds the OPE fit described in the text.}
\end{figure}

  We can also check the short distance behavior
of the correlation function in the instanton liquid. 
Instantons generate the same  $d=4$ operator 
in the OPE but the nature of the $d=6$ operator
is different. To leading order in the semi-classical
expansion there is no radiatively generated $\alpha_s
\langle\bar{q}q\rangle^2\log(x^2)x^6$ operator, but 
instead there is a non-singular $\langle\bar{q}q\rangle^2
x^6$ term. Such terms are dropped in standard OPE, but 
they are present in the correlation functions. The 
numerical value of this term is $(x/0.64\,{\rm fm})^2$,
close to the data and the OPE term.


4. We shall now focus our attention on the $V+A$ correlation function.
The unique feature of this function is that the full correlator
is close to the free field result for distances as large 
as 1 fm. This phenomenon was referred to as ``super-duality''
in \cite{Shuryak:1993kg}. 

 The instanton model reproduces this feature of the $V+A$ correlator.
We also notice that for small $x$ the deviation of the correlator
in the instanton model from free field behavior is small compared
to the perturbative $O(\alpha_s/\pi)$ correction. This opens the 
possibility of precision studies of the pQCD contribution. But 
before we do so, let us compare the correlation functions to
the OPE prediction
\bea
\frac{\Pi_V(x)+\Pi_A(x)}{2\Pi_0(x)} &=&  1+\frac{\alpha_s}{\pi}
  - \frac{1}{384}\langle g^2(G^a_{\mu\nu})^2 \rangle x^4 \nonumber \\
 & & \hspace{0.5cm}
  -\frac{2\pi^3}{81} \alpha_s(x)\langle{\bar{q}q}\rangle
   \log(x^2)x^6+\ldots 
\eea
Note that the perturbative correction is attractive, while the
power corrections of dimension $d=4$ and $d=6$ are repulsive.
Direct instantons also induce an $O(x^4)$ correction
$ 1 - \frac{\pi^2}{12} \left(\frac{N}{V}\right) x^4 + \ldots$ 
\cite{Andrei:1978xg,Dubovikov:1981bf,Nason:1994ak},
which is consistent with the OPE because in a dilute 
instanton liquid we have $\langle g^2G^2\rangle = 32\pi^2(N/V)$. 
This term can indeed be seen in the instanton calculation
and causes the correlator to drop below 1 at small $x$.


 It is possible to extract the value of $\Lambda_{QCD}$ together 
with the power corrections from the data. Because the
dimension 6 operator is relatively small we have fixed
it from a joint fit with the $V-A$ correlator. We find
\bea
\frac{\Pi_V(x)+\Pi_A(x)}{2\Pi_0(x)} &=& 1 + \frac{\alpha_s(x)}{\pi} 
-  \left(\frac{x}{1.52\,{\rm fm}}\right)^4
 \nonumber \\
& & \hspace{0.5cm} 
- \left(\frac{x}{0.85\,{\rm fm}}\right)^6 + \ldots .
\eea
The value of $\alpha_s(m_\tau)\simeq 0.35$ \cite{foot} is 
consistent with other determinations \cite{aleph1}, but the 
value of the gluon condensate term is smaller than the standard SVZ 
value \cite{SVZ}. In  fact, the data do not show any kind of ``dip'' 
and as soon as the $d=4$ power correction becomes comparable 
to the perturbative correction it is in dramatic disagreement 
with the data.  Unfortunately, it will be hard to improve on 
this situation even if high precision data that cover a larger 
range of invariant masses in the vector channel become available.
Within the range of validity of the OPE in the $V+A$ channel, 
$x\lsim 0.3 \, fm$, the power corrections are always small as 
compared to perturbative corrections. This makes it doubtful 
that one will ever be able to extract the value of the 
gluon condensate.

\begin{figure}[tbh]
\begin{center}
\vspace*{-1.2cm}
\epsfig{file=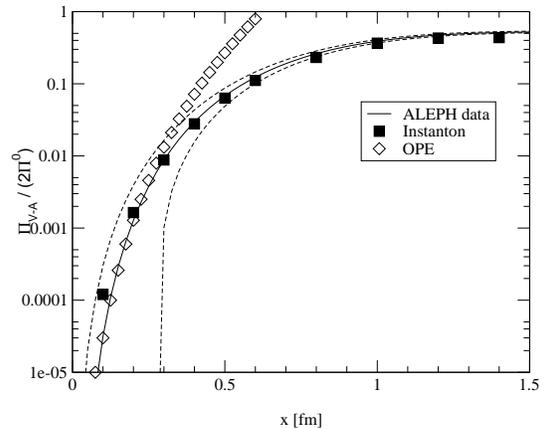,width=8cm,angle=-90}
\end{center}
\vspace*{-2.0cm}
\begin{center}
\epsfig{file=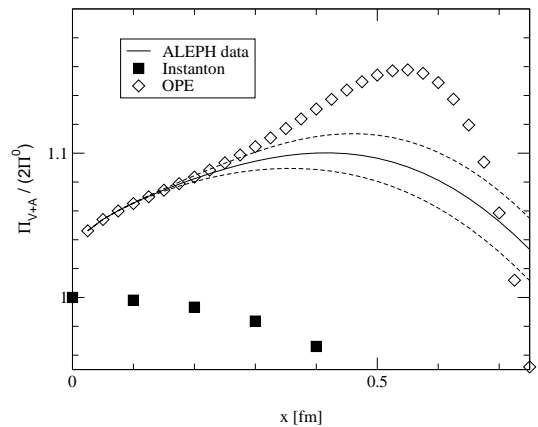,width=8cm,angle=-90}
\end{center}
\vspace*{-0.75cm}
\caption{\label{fig_cor2}
Same as in Fig. \ref{fig_cor1}, but with the $\Pi_V(x)-\Pi_A(x)$
correlator plotted on a logarithmic scale and $\Pi_V(x)+\Pi_A(x)$
shown in more detail.}
\end{figure}

5. Finally, we address the purely perturbative contribution 
to the $V+A$ correlation function, using the instanton calculation
as a representation of the non-perturbative part of the 
correlation function. This is supported by the fact that 
instantons provide a very accurate description of the $V-A$
correlator which is free of perturbative contributions.
The difference between the full correlation function and 
the instanton calculation is shown by the squares in Fig. 
\ref{fig_del}. For comparison, we also show the full correlation 
function with only the free field behavior subtracted. At short 
distance, there is no difference between the two curves, and both 
follow the first order perturbative result $\alpha_s(x)/\pi$.
At larger distances $\Pi_{V}(x)+\Pi_{A}(x)-2\Pi^0(x)$ 
starts to drop, but the non-perturbatively subtracted
correlator continues to grow. This behavior nicely shows
the running of $\alpha_s$ even at moderately large $x$
\cite{Girone:1996xb}. At $x\lsim 0.3$ fm the agreement
becomes even better if the two-loop contribution is 
added, but in this case the Landau pole is reached 
earlier. For this reason, the good agreement of the
data with the one-loop result even for large $x>0.3$ 
fm may be somewhat coincidental. The reason one is able 
to follow the pQCD behavior well outside the usual 
perturbative domain is the remarkable degree of 
cancellation among non-perturbative effects. Further high statistics
studies of this issue using lattice simulations would be very 
interesting.

\begin{figure}[tbh]
\begin{center}
\leavevmode
\vspace*{-1.2cm}
\epsfig{file=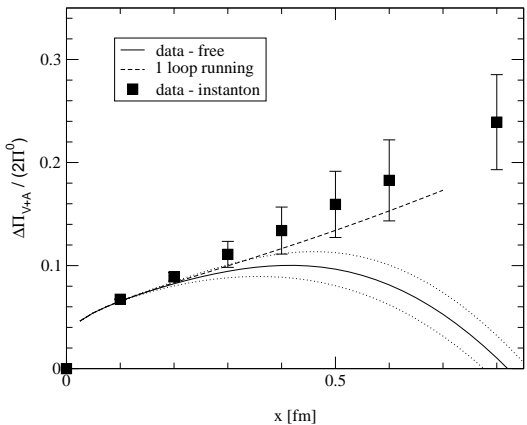,width=8cm,angle=-90}
\end{center}
\vspace*{-0.5cm}
\caption{\label{fig_del}
This figure shows an estimate of the perturbative part of the 
$V+A$ correlation function. The solid line is the measured
correlation function with the free field correlator subtracted.
The squares show the measured correlator with the instanton
contribution subtracted and the dashed line is the one-loop
prediction.}
\end{figure}

 6. In summary, we have used the  high statistics ALEPH data on 
hadronic $\tau$ decays to calculate Euclidean space 
correlation functions in the vector and axial-vector channel.
We focussed our discussion on the linear combinations $\Pi_V
(x)\pm\Pi_A(x)$. The combination $V-A$ receives no contribution
from perturbation theory and provides a clean probe for 
chiral symmetry breaking and the quark condensate. $V+A$,
on the other hand, allows for a study of perturbative QCD
and gluonic operators.

 We have compared the two correlation functions with the 
predictions of the random instanton liquid and the OPE.
The instanton model provides a very accurate description of 
the $V-A$ correlation function for all distances. In the 
$V+A$ channel the  instanton model, supplemented by
pQCD corrections with a running coupling constant,
also provides excellent description of the data for 
$x<1$ fm.

 The remarkable degree of cancellation of non-perturbative
effects in the $V+A$ channel provides a unique opportunity
to access perturbative corrections well beyond the usual 
pQCD domain. In the $V-A$ channel, on the other hand, there
is an opportunity to extract the dimension $d=6$ $\langle
\bar{q}q\rangle^2$ operator from the data. The result agrees
with the SVZ prediction, but the accuracy is limited by 
the largest invariant mass accessible in $\tau$ decays.
In addition to that, the instanton model suggests the presence
of a non-singular $d=6$ contribution of the same 
magnitude. Attempts to extract the $d=4$ gluon condensate 
operator from the $V+A$ channel fail because in the 
range of validity of the OPE the $d=4$ power correction 
remains a small correction to the pQCD contribution.

 We conclude that the range of validity of the OPE in the 
vector channels is quite small, $x\lsim 0.3$ fm. This means
that there is essentially no ``window'' in which both the OPE is
accurate and the correlation function is dominated by the 
ground state. Instantons, on the other hand, provide a 
quantitative tool at all distances. This is true even though 
the vector channels, because of the smallness of direct 
instanton effects, are generally considered to be the best 
system to study the OPE. 

{\bf Acknowledgments}:
We thank M. Shifman for valuable comments.
The work is partly supported by US DOE grant No. DE-FG02-88ER40388.

\end{narrowtext}

\begin{thebibliography}{99}
\bibitem{foot}
We have added higher order perturbative corrections not 
shown in the text.
\end{thebibliography}
\end{document}